# Strong terahertz radiation via rapid polarization reduction in photoinduced ionic-to-neutral transition of tetrathiafulvalene-*p*-chloranil

Yuto Kinoshita[1], Noriaki Kida[1], Yusuke Magasaki[1], Takeshi Morimoto[1], Tsubasa Terashige[2], Tatsuya Miyamoto[1], and Hiroshi Okamoto[1,2]

*[1]Department of Advanced Materials Science, The University of Tokyo, 5-1-5 Kashiwa-no-ha, Chiba 277-8561, Japan*

*[2]AIST-UTokyo Advanced Operando-Measurement Technology Open Innovation Laboratory, National Institute of Advanced Industrial Science and Technology, Chiba 277-8568, Japan*

Terahertz lights are usually generated through the optical rectification process within a femtosecond laser pulse in non-centrosymmetric materials. Here, we report a new generation mechanism of terahertz lights based upon a photoinduced phase transition (PIPT), in which an electronic structure is rapidly changed by a photoirradiation. When a ferroelectric organic molecular compound, tetrathiafulvalene-*p*-chloranil, is excited by a femtosecond laser pulse, the ionic-to-neutral transition is driven and simultaneously a strong terahertz radiation is produced. By analyzing the terahertz electric-field waveforms and their dependence on the polarization direction of the incident laser pulse, we demonstrate that the terahertz radiation originates from the ultrafast decrease of the spontaneous polarization in the photoinduced ionic-to-neutral transition. The efficiency of the observed terahertz radiation via the PIPT mechanism is found to be much higher than that via the optical rectification in the same material and in a typical terahertz emitter, ZnTe.

Recent developments in the generation and detection techniques of a terahertz electromagnetic wave or a terahertz pulse have opened a new possibility for their various applications such as imaging, sensing, and spectroscopy [1,2]. The exploration of a strong terahertz-radiation source becomes more and more important in developing terahertz science and technology. Within the classical electromagnetism, a generation of a terahertz pulse can be understood as the electric-dipole radiation, which is given by

$$\boldsymbol{E}_{\text{THz}} \propto \frac{\partial^2}{\partial t^2}\boldsymbol{P}. \qquad (1)$$

Here, $\boldsymbol{E}_{\text{THz}}$ is the electric field of the generated terahertz pulse and $\boldsymbol{P}$ is the time-dependent polarization. In this framework, a most popular mechanism to generate a terahertz pulse is an optical rectification (OR) [3] or equivalently a difference frequency generation (DFG) within a femtosecond laser pulse via the second-order optical nonlinearity [4]. The induced nonlinear polarization is expressed by $P_i^{(2)} = \epsilon_0 \chi_{ijk}^{(2)} E_j^\omega E_k^{\omega*}$, in which $\chi_{ijk}^{(2)}$ is the second-order nonlinear susceptibility, $\epsilon_0$ is the vacuum permittivity, and $E^\omega$ is the light electric field [4]. As a femtosecond laser source, an output of a titanium-sapphire (TS) laser (a wavelength of 800 nm and a photon energy of 1.55 eV) with a temporal width of 100 fs, is frequently used. That pulse has a finite spectral width of ~5 THz, and the DFG can occur within the pulse from the nonlinear polarization $P^{(2)}$. This results in the generation of a broadband terahertz pulse with the central frequency of ~1 THz, the electric-field waveform of which is almost proportional to the second time-derivative of the envelope of the incident femtosecond laser pulse [1,5,6]. Using this process in a transparent media with no space inversion symmetry, e.g., inorganic crystals [1,7,8] such as ZnTe and GaP and organic crystals [9,10] such as 4-N,N-dimethylamino-4'-N'-methyl stilbazolium tosylate (DAST), a stable terahertz pulse can be obtained. In the OR process, however, it is difficult to enhance the electric-field amplitude of a generated terahertz pulse, since two-photon absorption of an incident femtosecond laser pulse necessarily occurs, which limits the pulse fluence introduced to the nonlinear optical crystal. To overcome this difficulty, the pulse-front-tilting method using TS regenerative amplifier (TSRA) and a wide-gap nonlinear optical crystal LiNbO$_3$ [2], and another DFG method [11,12] combining an organic nonlinear optical crystal, 4-$N$, $N$-dimethylamino-4′-$N$′-methyl-stilbazolium 2,4,6-trimethylbenzenesulfonate (DSTMS) and a near-IR femtosecond laser pulse have been proposed, although they require a special optical set-up and a strong near-IR laser source different from TSRA, respectively.

In this paper, we report a new generation mechanism of a strong terahertz pulse using a photoinduced phase transition (PIPT), in which a ferroelectric phase is melted by the irradiation with a femtosecond laser pulse. In the PIPT, a macroscopic polarization is rapidly decreased, which can induce a strong terahertz radiation via Eq. (1). The studied material is a mixed-stacked organic molecular compound, tetrathiafulvalene-$p$-chloranil (TTF-CA), which is one of the most famous compounds showing PIPTs [13-17]. We demonstrate that the efficiency of the terahertz radiation via the PIPT in TTF-CA is much higher than that via the OR in a typical terahertz emitter, ZnTe.

Figure 1(a) shows the crystal structure of TTF-CA. In this compound, donor ($D$) TTF molecules and accepter ($A$) CA molecules stack alternately along the $a$-axis, resulting in the formation of the quasi-one-dimensional electronic structure. The electronic state of TTF-CA is characterized by the degree of charge-transfer (CT), $\rho$, from $A$ to $D$ molecules [18]. At room temperature, TTF-CA is a van-der-Waals neutral crystal [Fig. 1(b)] with $\rho_N$ ~0.3. With decreasing temperature, it undergoes the neutral($N$)-to-ionic($I$) phase transition at $T_c$ = 81 K and $\rho$ increases ($\rho_I$~0.6) [19-21]. The $I$ phase is stabilized by the energy gain because of the long-range Coulomb attractive interaction or the Madelung potential. The lattice contraction caused by the decrease of temperature enhances the Madelung potential, which drives the *NI* transition [19,20,22]. In the *I* phase, each molecule has a spin 1/2. As a result, *DA* molecules are dimerized as shown by the ellipses in Fig. 1(c) because of the spin-Peierls-like instability. The dimeric molecular displacements are three dimensionally ordered, breaking the space inversion symmetry along the *a*-axis [20,22]. The recent measurement of the polarization–electric-field (*P-E*) characteristic demonstrates that TTF-CA shows a spontaneous polarization $P_S$ along the *a*-axis with a hysteresis loop in the *I* phase [23]. The X-ray scattering experiments under electric fields and the theoretical simulations revealed that the direction of $P_S$ is anti-parallel (parallel) to the displacement direction of $D^{+\rho}$ ($A^{-\rho}$) molecules as shown by the yellow arrow in Fig 1(c). These results indicate that $P_S$ is not caused by the displacement of the ionic molecules [23] but originates from the collective intradimer electron transfers, which occur across the *NI* transition. In fact, the magnitude of $P_S$ is 6.3 $\mu C/cm^2$, which is about 20 times larger than that estimated from the point-charge model [23-25]. This type of ferroelectricity is called electronic ferroelectricity.

This *NI* transition can also be induced by a photoirradiation. When TTF-CA is irradiated with a femtosecond laser pulse below $T_c$, the *I* state is converted to *N* state [14,16,26]. The detailed analyses of the pump-probe reflection spectroscopy suggested that when the excitation photon energy is 0.65 eV, which corresponds to the lowest CT transition, an *N* domain consisting of ~8$D^0A^0$ pairs is produced per photon as schematically shown in Fig. 1(d). When the excitation photon density is increased, a large part of the *I* state can be converted to the *N* state, which is called a photoinduced *I*-to-*N* transition. The initial *I*-to-*N* conversion processes occurs within 20 fs and are considered to be purely electronic in nature [26]. Such an *I*-to-*N* conversion decreases the ferroelectric polarization as ascertained by the optical-pump second-harmonic-generation-probe measurements [27]. This makes us expect the strong terahertz radiation during the photoinduced *I*-to-*N* transition.

(001)-oriented single crystals of TTF-CA were grown by the co-sublimation method, details of which was reported elsewhere [16]. The flesh (001)-surface with the thickness of ~250 μm was obtained by cleaving an as-grown single crystal. The sample was cooled with a rate of 0.33 K/min. to avoid a clacking of the crystal.

For the terahertz radiation experiments, an optical parametric amplifier (OPA) excited by an output of a TSRA with the central wavelength of 800 nm, the repetition rate of 1 kHz, and the temporal width of 100 fs was used. The excitation pulse is set at 0.65 eV corresponding to the CT transition, and focused on the *ab* plane of TTF-CA in the normal incidence. The penetration depth of the 0.65-eV light is ~40 nm at 15 K [16], which is much shorter than the typical thickness of used single crystals (~150-300 μm), and only the surface region of the crystal is photoexcited. To avoid additional absorptions of the terahertz radiation by infrared-active phonons [28-31], we use the reflection configuration as illustrated in Fig. 2(a) [32] instead of the transmission configuration generally used for terahertz-radiation measurements. Electric-filed waveforms of generated terahertz pulses along the *a*-axis are measured by a standard electro-optic (EO) sampling with a 1-mm-thick (110) ZnTe crystal. We confirmed that the emitted terahertz pulses are polarized along the *a*-axis. By using terahertz radiation imaging [33,34] at 15 K, we confirmed that the TTF-CA crystal consists of a single ferroelectric domain. The result of the domain imaging is shown in the Supplemental Material S1 [35].

A typical electric-field waveform of a terahertz pulse generated by the 0.65-eV excitation at 15 K is shown in Fig. 2(b) (the blue line). The excitation photon density per unit area was $1.6 \times 10^{15}$ photon/cm$^2$. Within the penetration depth of the excitation light (~40 nm), the excitation density is estimated to be 0.11 photon/DA pair. The incident electric field of a femtosecond laser pulse is set parallel to the $a$-axis ($E_{\text{ex}}^{\omega} \parallel a$). In our experiments, we cannot determine the time origin (0 ps), which should be defined as the time when the incident laser pulse reaches the sample. Instead, we set the time origin at the time when the absolute value of the terahertz electric field $|E_{\text{THz}}(t)|$ for $E_{\text{ex}}^{\omega} \parallel b$ ($E_{\text{ex}}^{\omega} \perp a$) is the maximum, as discussed later. The generated terahertz pulse consists of a single-cycle component around 0 ps and the subsequent oscillatory component.

The blue line in Fig. 2(c) displays the Fourier power spectrum of the single-cycle component of the data from -2.0 ps to 0.8 ps indicated by an arrow in Fig. 2(b). It shows the peak structure at ~0.98 THz and expands to ~2.20 THz. The orange line in the same figure shows the Fourier power spectrum of the oscillatory component in the range of 0.8-8 ps, which exhibits a peak structure at 1.60 THz (~53 cm$^{-1}$). The tiny dip at ~1.37 THz is due to the absorption of quartz used for a window of the cryostat, which is detailed in the Supplemental Material S2 [35]. In order to characterize the magnitude of the terahertz electric field in TTF-CA, we measured the terahertz electric-field waveform in a (110)-oriented ZnTe crystal in the same experimental setup. ZnTe is known to show a strong terahertz radiation via the optical rectification (OR) mechanism. The results reveal that the maximum amplitude (intensity) of the terahertz electric field in TTF-CA is approximately 25 (600) times as large as that in ZnTe. The result is detailed in the Supplemental Material S3 [35].

To reveal the terahertz-radiation mechanism in TTF-CA, we measured how the terahertz electric-field waveform of the terahertz pulse depends on the direction of the light electric field $E_{\text{ex}}^{\omega}$ of the incident femtosecond pulse. Figures 3(a) and (b) show the waveforms of terahertz electric fields with $E_{\text{THz}}(t) \parallel a$ at 15 K for $E_{\text{ex}}^{\omega} \parallel a$ and $E_{\text{ex}}^{\omega} \parallel b$ ($E_{\text{ex}}^{\omega} \perp a$), respectively. The excitation photon density was 1.6×10$^{15}$ photon/cm$^2$ in common. The time origin is set at the time when $|E_{\text{THz}}(t)|$ for $E_{\text{ex}}^{\omega} \parallel b$ is the maximum.

The maximum value of $|E_{\mathrm{THz}}(t)|$ for $E_{\mathrm{ex}}^{\omega} \parallel b$ is about 1/2.5 of that for $E_{\mathrm{ex}}^{\omega} \parallel a$. Noticeably, the observed waveform for $E_{\mathrm{ex}}^{\omega} \parallel b$ [Fig. 3(b)] is considerably different from that for $E_{\mathrm{ex}}^{\omega} \parallel a$ [Fig. 3(a)]. As guided by dotted vertical lines at 0 ps, $|E_{\mathrm{THz}}(0)|$ is nearly equal to zero for $E_{\mathrm{ex}}^{\omega} \parallel a$, when $|E_{\mathrm{THz}}(0)|$ is the maximum for $E_{\mathrm{ex}}^{\omega} \parallel b$. The Fourier power spectrum obtained using the waveform from -2.0 ps to 0.8 ps for $E_{\mathrm{ex}}^{\omega} \parallel b$ shows the peak at ~0.60 THz [as shown later in Fig. 4(b)], which is slightly lower than the peak (0.98 THz) for $E_{\mathrm{ex}}^{\omega} \parallel a$ [Fig. 2(c)].

The observed polarization dependence of the electric-field waveforms suggests that two different mechanisms of terahertz radiations exist in TTF-CA. Since the sample is transparent for the 0.65-eV light with $E_{\mathrm{ex}}^{\omega} \parallel b$, the terahertz radiation for $E_{\mathrm{ex}}^{\omega} \parallel b$ can be ascribed to the OR mechanism. For 0.65-eV light with $E_{\mathrm{ex}}^{\omega} \parallel a$, the PIPT occurs. Therefore, it is natural to consider that the terahertz radiation for $E_{\mathrm{ex}}^{\omega} \parallel a$ is attributed to the decrease of the ferroelectric polarization in the photoinduced *I*-to-*N* transition (the PIPT mechanism).

To ascertain this interpretation, we next simulated the terahertz electric-field waveform via the PIPT mechanism. In the simulations, it should be considered that an electric-field waveform experimentally obtained is affected by the frequency dependence of the detection sensitivity in the EO sampling. In our study, we correct this effect by considering the electric-field waveform for $E_{\mathrm{ex}}^{\omega} \parallel b$, the generation mechanism of which is the OR. By comparing the experimental electric-field waveform with the theoretical one, we can determine a response function $\beta(\omega)$ of our measurement system. The derivation of $\beta(\omega)$ and the simulation of the terahertz radiation via the PIPT mechanism are detailed in the Supplementary Material S4 [35]. If the OR mechanism is dominant, the ideal amplitude spectrum of the terahertz electric field is proportional to the Fourier power spectrum of the second time-derivative of $P^{(2)}$, which is given by $E_{\mathrm{THz}}^{\mathrm{OR}}(\omega) \propto$ $(i\omega)^2 I(\omega)$, where $I(\omega)$ is a Fourier power spectrum of an envelope function of the incident femtosecond laser pulse [5,6]. On the other hand, if the PIPT is dominant, the ideal amplitude spectrum of the electric field is given by $E_{\text{THz}}^{\text{PIPT}}(\omega) \propto (i\omega)^2 I(\omega) \frac{1}{i\omega+a}$, which includes a relaxation function with a time constant $1/a$. By using $\beta(\omega)$, we can simulate the terahertz electric-field waveform originating from the PIPT, which is shown by the green line in Fig. 3(a). The simulated waveform for the PIPT successfully reproduces the characteristic feature that $|E_{\text{THz}}(0)|$ is zero for $E_{\text{ex}}^{\omega} \parallel a$ [Fig. 3(a)] when $|E_{\text{THz}}(0)|$ is maxima for $E_{\text{ex}}^{\omega} \parallel b$ [Fig. 3(b)].

The coherent oscillation [the orange line in Fig. 2(a)] is observed only for $E_{\text{ex}}^{\omega} \parallel a$, which is detailed in the Supplementary Material S5 [35], and can also be related to the PIPT. A similar oscillation was observed in the reflectivity change at the intramolecular transition band of TTF in the visible region during the photoinduced *I*-to-*N* transition [16,26,36] and assigned to the coherent oscillation corresponding to the release of the dimerization in the photogenerated *N* domains. This coherent oscillation of the dimeric mode slightly modulates the degree of CT $\rho_N$ in the photogenerated *N* states as $\rho_N \pm \delta\rho$, which is schematically shown in Fig. 2(d). Such a charge modulation gives rise to the terahertz radiation with the same frequency.

In order to further discuss the terahertz-radiation mechanism in TTF-CA, we measured the electric-field waveforms of the terahertz radiations by changing the polarization angle $\theta$ of the incident pulse with a step of $5°$. The results are shown in Fig. 4(a), and their Fourier power spectra are plotted in Fig. 4(b). By changing $\theta$ from $0°$ to $90°$, the central frequency of the Fourier power spectrum is shifted to the lower frequency, while the peak intensity decreases. In order to see this tendency more clearly, we plotted the magnitudes of the terahertz radiation intensities at selected frequencies as a function of $\theta$ (0° to 180°) in Fig. 4(c). The intensity reaches the maxima at $\theta = 0°$ and $\theta = 180°$ ($E_{\text{ex}}^{\omega} \parallel a$) and the intensity below 1.5 THz is finite at $\theta = 90°$ ($E_{\text{ex}}^{\omega} \parallel b$).

We can also simulate the observed $\theta$ dependence of the terahertz radiation intensity. First, we discuss the component for the PIPT. As mentioned above, the PIPT occurs only for $E_{\text{ex}}^{\omega} \parallel a$ in TTF-CA [16]. In this case, the terahertz radiation intensity along the $a$-axis, $I_a^{\text{PIPT}}$, is expressed as

$$I_a^{\text{PIPT}} = |E_a^{\text{PIPT}}|^2 \propto d_{\text{PIPT}}^2 \cos^4\theta \, E_0^4. \tag{2}$$

Here, $d_{\text{PIPT}}$ is the coefficient characterizing the terahertz radiation efficiency and $E_0$ is the electric field of the incident light. The fitting curves based on Eq. (2) cannot reproduce the finite intensity at $\theta = 90°$ ($E_{\text{ex}}^{\omega} \parallel b$), which is shown in Fig. S8 in the Supplementary Material S6 [35]. To discuss the polarization angle dependence of the terahertz radiation via the OR mechanism, the symmetry analysis is necessary. TTF-CA in the $I$ phase has the space group $m$ and the $d$-tenor can have finite values of $d_{11}$, $d_{12}$, $d_{13}$, and $d_{16}$ in our experimental configuration [the inset of Fig. 4(a)] [4]. The intensity of the terahertz radiation polarized along the $a$-axis due to the OR, $I_a^{\text{OR}}$, is expressed as

$$I_a^{\text{OR}} = |E_a^{\text{OR}}|^2 \propto |(d_{11}\cos^2\theta + d_{12}\sin^2\theta)|^2 E_0^4. \tag{3}$$

Using the fitting curves based on Eq. (3) or the combination of Eqs. (2) and (3), however, we could not completely reproduce the experimental data (see Fig. S8). We next introduced the phase difference $\phi$ between Eqs. (2) and (3), which gives the expression of the radiation intensity as

$$I_a = |E_a^{\text{OR}} + E_a^{\text{PIPT}}\exp(i\phi)|^2. \tag{4}$$

Using this formula with $\phi = 90°$ and $d_{11}/d_{12} = 3.05$, the observed $\theta$ dependence of the radiation intensity can be well reproduced as shown by thin solid lines in Fig. 4(c). Thus, we conclude that the terahertz radiation in TTF-CA can be explained by the combination of the OR and PIPT mechanisms. This conclusion is also supported by the incident laser-power dependence of the terahertz radiation intensity (Supplementary Material S7 [35]). In Fig. 4(d), we show the spectrum of the square of the $d_{\mathrm{PIPT}}$ ($d_{11}$) value characterizing the terahertz radiation efficiency by the PIPT (OR) mechanism by blue (red) circles. The magnitude of $d_{\mathrm{PIPT}}^2$ is approximately one order larger than that of $d_{11}^2$ dominating the OR. This indicates that the rapid polarization reduction via the photoinduced *I*-to-*N* transition in TTF-CA is a very efficient mechanism as a light-induced terahertz radiation.

In summary, we observed the strong terahertz radiation via the reduction of the ferroelectric polarization in the photoinduced ionic-to-neutral transition of TTF-CA. The intensity of the terahertz electric field is approximately 10 times as large as that caused by the optical-rectification in the same sample and approximately 600 times as large as that in ZnTe. Thus, our result provides a new way for obtaining the strong terahertz radiation by the photoinduced-phase-transition mechanism, which will be able to be used for various applications of terahertz radiations.

This work was partly supported by a Grant-in-Aid by MEXT (No. JP25247058, JP25600072, JP15K13330, and JP18H01858) and CREST (JPMJCR1661), Japan Science and Technology Agency. Y. Kinoshita and T. Morimoto were supported by Japan Society for the Promotion of Science (JSPS) through Program for Leading Graduate Schools (MERIT) and JSPS Research Fellowships for Young Scientists.

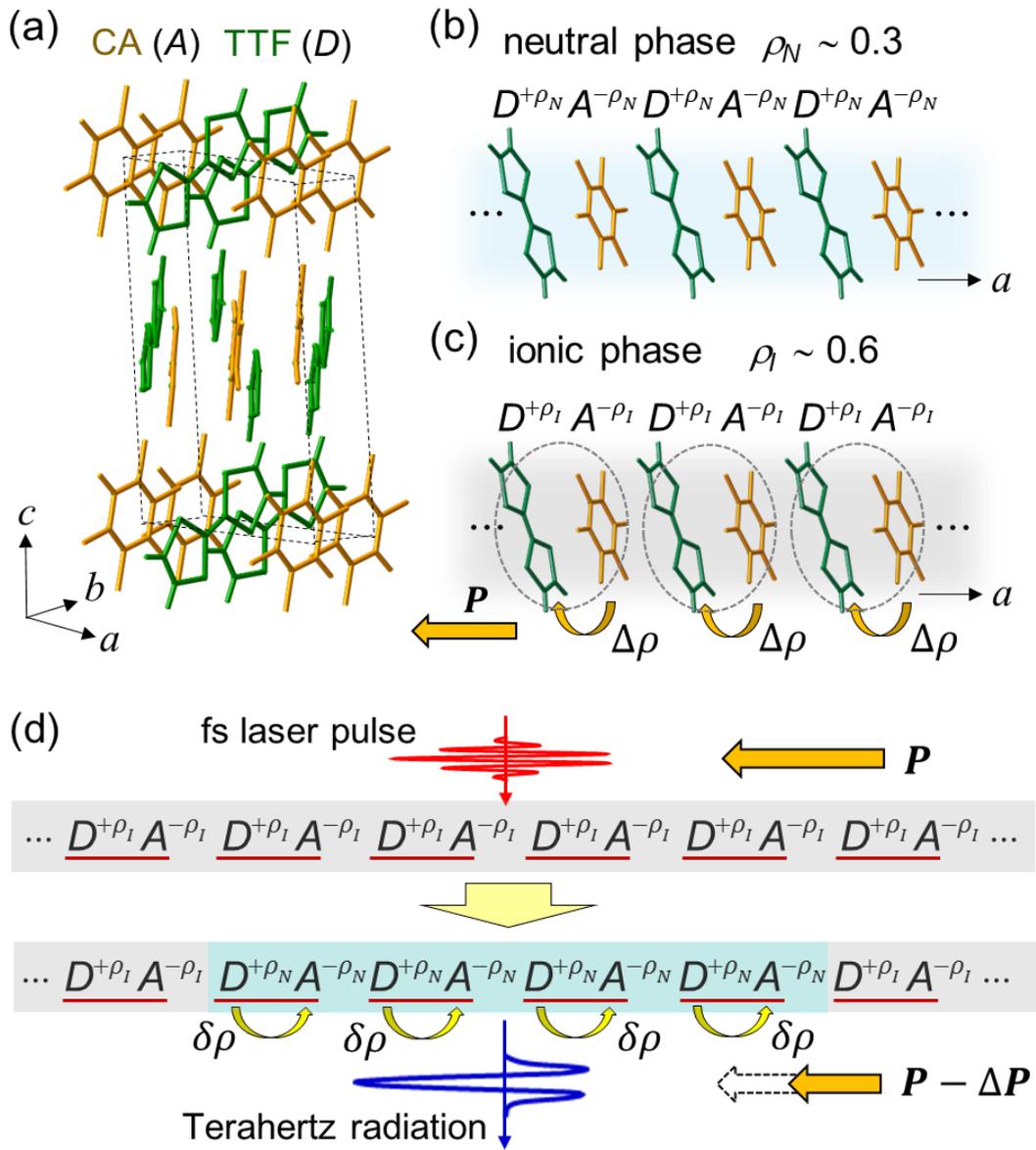

FIG. 1. (a) Crystal structure of TTF-CA. (b, c) Molecular stacks in (b) neutral and (c) ionic phases. (d) Schematic of the ionic-to-neutral transition induced by a femtosecond laser pulse and terahertz radiation via the reduction of the polarization.

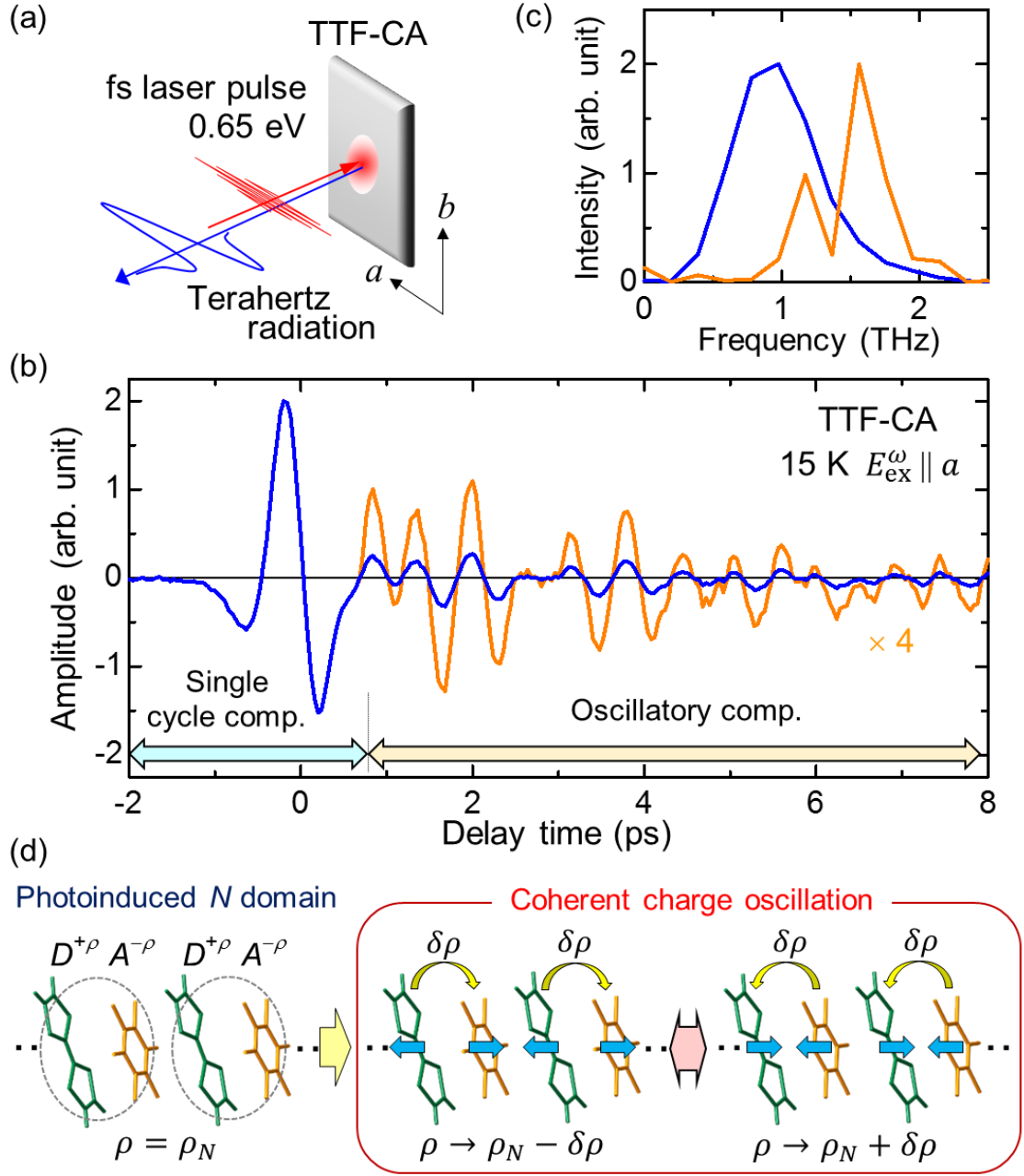

FIG. 2. (a) Schematic of the terahertz radiation measurement on TTF-CA in the reflection configuration. (b) Terahertz electric-field waveform for $E^{\omega}_{ex} \parallel a$, measured at 15 K. (c) Normalized Fourier power spectra of the data shown in (b); the blue line shows the single cycle component at -2 to 0.8 ps and the orange line shows the oscillatory component at 0.8 to 8 ps. (d) Schematic of a coherent charge oscillation in a photoinduced $N$ domain originating from the release of the dimerization.

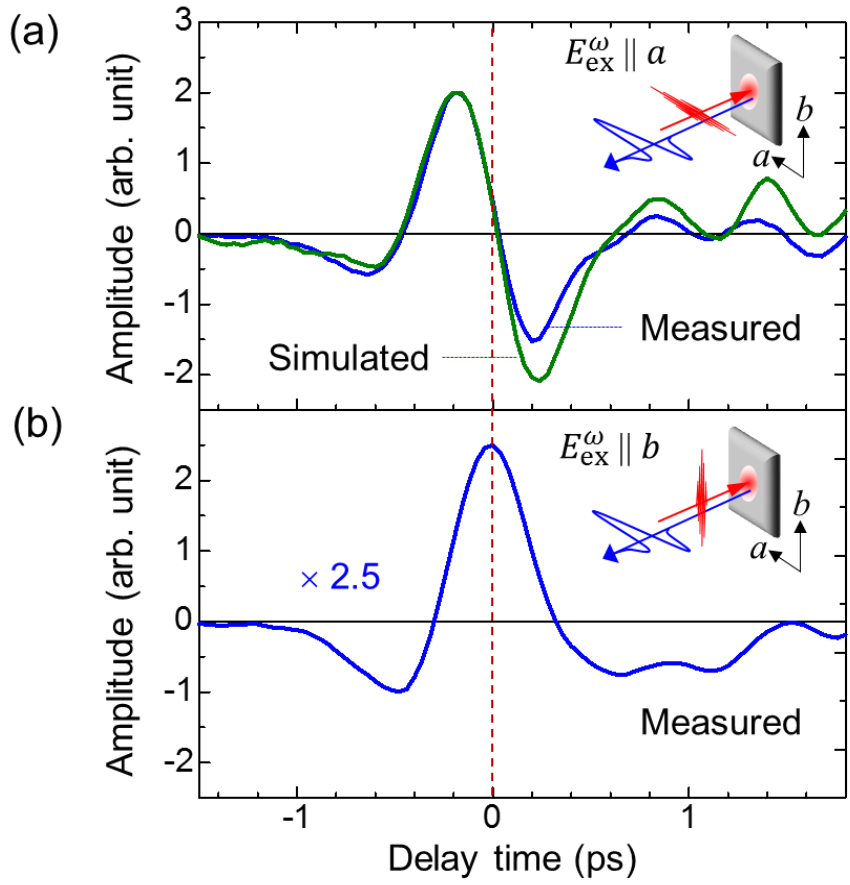

FIG. 3. The measured terahertz electric-field waveforms for (a) $E^{\omega}_{ex} \parallel a$ and (b) $E^{\omega}_{ex} \parallel b$, respectively. The simulated waveform is also shown by the green line in (a).

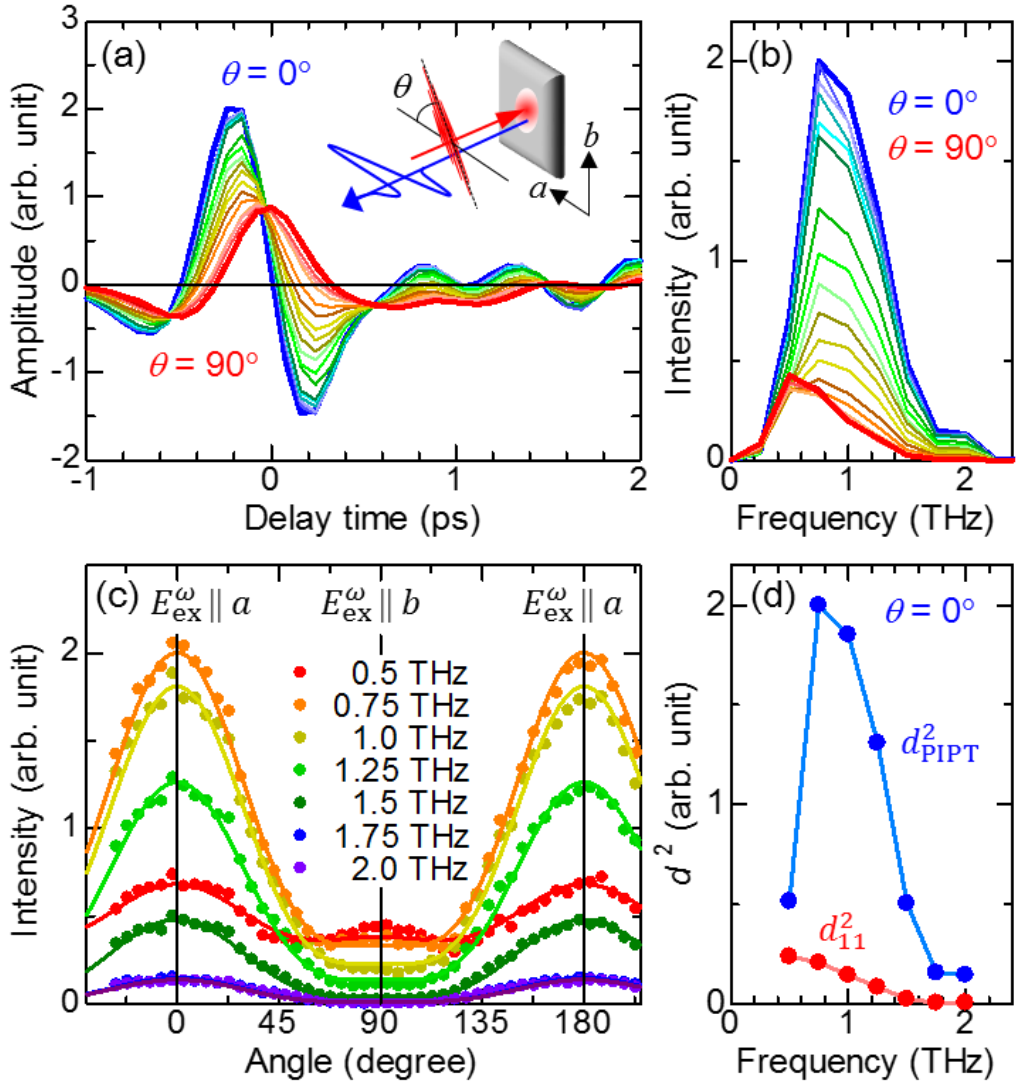

FIG. 4. Light-polarization angle dependence of (a) terahertz electric-field waveforms and (b) their Fourier power spectra. (c) Intensities of Fourier power spectra in (b) at various frequencies as a function of the angle $\theta$. (d) The frequency dependence of the square of the $d$-tensor originating from the optical rectification ($d_{11}^2$) and photoinduced phase transition ($d_{PIPT}^2$).

19